\documentclass[pre,showpacs,preprint,tightenlines,nofootinbib,eqsecnum]{revtex4}

\usepackage{graphicx,bm,amsfonts}
\usepackage[tbtags]{amsmath}

\newcommand{\ij}{i\kern -0.08em j}
\newcommand{\half}{{\textstyle\frac{1}{2}}}
\def\hn{\mskip-0.5\thinmuskip}
\def\hp{\mskip0.5\thinmuskip}

\def\re{\mathop{\rm Re}\nolimits}
\def\diag{\mathop{\rm diag}\nolimits}
\def\beq{\begin{equation}}
\def\eeq{\end{equation}}
\def\eeql#1{\label{#1} \end{equation}}

\newcommand{\ua}{\uparrow}

\newcommand{\To}{\rightarrow}
\newcommand{\M}{\hphantom{-}}
\newcommand{\ir}{\mathrm{i}}
\newcommand{\dr}{\mathrm{d}}
\newcommand{\e}{\mathrm{e}}
\newcommand{\pt}{\partial}
\newcommand{\lm}{\lambda}
\newcommand{\Lm}{\Lambda}
\newcommand{\Oc}{\mathcal{O}}
\newcommand{\nl}{\hphantom{0}}

\begin{document}

\title{Hamiltonian for coupled flux qubits}
\author{Alec \surname{Maassen van den Brink}}
\email{alec@dwavesys.com}
\affiliation{D-Wave Systems Inc., 320-1985 West Broadway, Vancouver, B.C., V6J 4Y3 Canada}

\date{\today}

\begin{abstract}
An effective Hamiltonian is derived for two coupled three-Josephson-junction (3JJ) qubits. This is not quite trivial, for the customary ``free'' 3JJ Hamiltonian is written in the limit of zero inductance~$L$. Neglecting the self-flux is already dubious for one qubit when it comes to readout, and becomes untenable when discussing inductive coupling. First, inductance effects are analyzed for a single qubit. For small $L$, the self-flux is a ``fast variable'' which can be eliminated adiabatically. However, the commonly used junction phases are \emph{not} appropriate ``slow variables'', and instead one introduces degrees of freedom which are decoupled from the loop current to leading order. In the quantum case, the zero-point fluctuations ($LC$ oscillations) in the loop current diverge as $L\To0$. While their effect thus formally dominates over the classical self-flux, it merely renormalizes the Josephson couplings of the effective (two-phase) theory.

In the coupled case, the strong zero-point fluctuations render the full (six-phase) wave function significantly entangled in leading order. However, in going to the four-phase theory, this uncontrollable entanglement is integrated out completely, leaving a computationally usable mutual-inductance term of the expected form as the effective interaction.
\end{abstract}

\maketitle

\section{Introduction}

A commonly considered flux qubit consists of a superconducting loop with three Josephson junctions---a 3JJ qubit~\cite{orlando}. One readily writes down its Hamiltonian,\footnote{As always, $\phi_i$ in (\ref{H3phi}) are properly gauge-invariant ``phases'', defined as line integrals along the SQUID loop. Thus, there is no reason for $H$ to be $2\pi$-periodic even though the Josephson term is, and indeed $\phi_i$'s differing by multiples of $2\pi$ correspond to physically distinguishable magnetic field configurations.}
\beq
  H=\sum_{i=1}^3\left[\frac{Q_i^2}{2C_i}-E_i\cos\phi_i\right]
  +\frac{(\phi_1{+}\phi_2{+}\phi_3{-}\phi_\mathrm{x})^2}{8e^2L}\;,
\eeql{H3phi}
in units with $\hbar=1$, and with the external flux bias $\Phi_\mathrm{x}$ given in phase units,
\beq
  \phi_\mathrm{x}=2e\Phi_\mathrm{x}\;.
\eeql{Phi-phi}
As long as one neglects gate capacitors etc., the $C_i$ are simply the junction capacitances: for finite inductance~$L$, the redistribution of charges following a tunneling event, often accounted for by an effective capacitance matrix, is not instantaneous, i.e., all three junction charges and phases are independent dynamical variables.\footnote{A nontrivial capacitance matrix will be essential in~(\ref{H3JJ}). Note also that, for this pure series circuit, $H$~does not depend on how the inductance is distributed along the loop.\label{series}}

The essence of the 3JJ design is to introduce bistability without relying on magnetic energy, so that the SQUID loop can be kept small. For $L\To0$, the last term in (\ref{H3phi}) mainly implements the constraint $\phi_3=\phi_\mathrm{x}-\phi_1-\phi_2$. This leads to an effective two-phase theory,
\begin{align}
  H_\text{3JJ}&=\half\vec{Q}^\mathrm{T}\mathsf{C}^{-1}\vec{Q}
  -E_1\cos\phi_1-E_2\cos\phi_2-E_3\cos(\phi_\mathrm{x}{-}\phi_1{-}\phi_2)\;,\label{H3JJ}\\
  \mathsf{C}&=\begin{pmatrix} C_1+C_3 & C_3 \\ C_3 & C_2+C_3 \end{pmatrix}\;,
\end{align}
with $\vec{Q}=(Q_1,Q_2)^\mathrm{T}$ (cf.\ (4) in \cite{orlando}, bearing in mind that $\phi_1$ and $\phi_2$ have the opposite relative sign there).

When scaling to circuits with more than one qubit, the question arises how to generalize the effective Hamiltonian (\ref{H3JJ}). Unfortunately, repeating the above reduction merely yields uncoupled 3JJ Hamiltonians. Clearly, by neglecting self-inductances entirely one has missed the effect of mutual inductance (which can never be larger), responsible for qubit--qubit interaction and ultimately entanglement. This can be turned into an advantage: most of the analysis can be done by studying \emph{one} qubit to higher order in $L$ than usual, after which two qubits require only a straightforward generalization.

This is taken up in Section~\ref{class} on the level of classical Kirchhoff (circuit) equations. Section~\ref{quant} is devoted to the quantum case. Rather than merely confirming the classical result or resolving operator-ordering ambiguities, one finds an additional physical effect: zero-point fluctuations in the loop current wash out (renormalize) the effective Josephson couplings. The results are verified numerically, and critically compared to a previous attempt~\cite{C&O}. Section~\ref{couple} contains the generalization to two qubits; it is seen that the four-phase wave function is entangled only in considerably higher order in the inductance than the six-phase one. This integrating-out of uncontrollable entanglement may indicate that the effective theory is not merely computationally convenient, but also physically appropriate. Some concluding remarks are made in Section~\ref{disc}.

For small $L$, self-fluxes vanish but persistent currents remain finite, so that the latter are often more convenient. When studying dynamics etc., one needs the current \emph{operator} (as opposed to, say, the ground-state expectation~\cite{greenberg}), which is not readily available in the growing literature on the 3JJ qubit. Preliminary to the $L$-expansion proper, it will now be derived quantum-mechanically in both the three- and two-phase theories, which has some independent interest. In fact, one merely needs to add the capacitive and Josephson contributions. Starting from (\ref{H3phi}), one has\footnote{When choosing conventions, one needs a minus sign somewhere for the final (\ref{I3}) with (\ref{Phi-phi}) to come out as $\Phi-\Phi_\mathrm{x}=LI$.}
\beq
  I_i=-I_{\mathrm{c}i}\sin\phi_i-\dot{Q}_i\;,
\eeq
with $I_{\mathrm{c}i}\equiv 2eE_i$ and
\beq
  Q_i=\frac{2e}{\ir}\frac{\pt}{\pt\phi_i}\;.
\eeql{Qquant}
Then,
\begin{align}
  \dot{Q}_i&=[H,\ir Q_i]=\frac{\phi_\mathrm{x}-\sum_j\phi_j}{2eL}
    -I_{\mathrm{c}i}\sin\phi_i\quad\Rightarrow\\
  I_i&=\frac{\sum_j\phi_j-\phi_\mathrm{x}}{2eL}\label{I3}\;.
\end{align}
That is, $I_i=I$ independent of $i$ as expected; contrast, e.g., arbitrarily picking the Josephson current through one of the junctions.

For $L\To0$, (\ref{I3}) tends to $\frac{0}{0}$ and a separate derivation from (\ref{H3JJ}) is needed. In this case, however, the canonical variable $Q_1$ is not simply the charge on capacitor~1 (indeed, two $Q$'s have to account for three $C$'s); to find the latter, we set $\mathsf{C}^{-1}=\bigl(\begin{smallmatrix} p & r \\ r & q \end{smallmatrix}\bigr)$ and evaluate
\beq
  \dot{\phi}_1=\ir[\half pQ_1^2{+}rQ_1Q_2,\phi_1]=2e(pQ_1+rQ_2)\;.\label{dot-phi}
\eeq
Hence,
\beq\begin{split}
  I_1^{(0)}&=-\frac{C_1}{2e}\ddot{\phi}_1-I_{\mathrm{c}1}\sin\phi_1\\
  &=(C_1p{-}1)I_{\mathrm{c}1}\sin\phi_1+C_1rI_{\mathrm{c}2}\sin\phi_2
    -C_1(p{+}r)I_{\mathrm{c}3}\sin(\phi_\mathrm{x}{-}\phi_1{-}\phi_2)\\
  &=-C\biggl[\frac{I_{\mathrm{c}1}}{C_1}\sin\phi_1+\frac{I_{\mathrm{c}2}}{C_2}\sin\phi_2
    +\frac{I_{\mathrm{c}3}}{C_3}\sin(\phi_\mathrm{x}{-}\phi_1{-}\phi_2)\biggr]\;,
\end{split}\eeql{I0}
an appealing and manifestly symmetric form $I_1^{(0)}=I^{(0)}_{\vphantom{1}}$, involving the loop's series capacitance $C^{-1}=C_1^{-1}+C_2^{-1}+C_3^{-1}$. Interestingly, the junction area~$\kappa_i$ cancels from~(\ref{I0}): if, as is often assumed, $I_{\mathrm{c}i}\propto C_i\propto\kappa_i$, then the three sines have equal prefactors.

\section{Classical analysis}
\label{class}

The Hamilton equations
\beq
  \dot{\phi}_i=2e\frac{\pt H}{\pt Q_i}\;,\qquad
  \dot{Q}_i=-2e\frac{\pt H}{\pt\phi_i}
\eeql{Hdyn}
yield the classical dynamics of the system~(\ref{H3phi}).\footnote{Since superconductivity is not a ``classical'' phenomenon, the term is a bit tenuous. Indeed, if $H$ is written in terms of phases (fluxes), the inductive (Josephson) terms would involve $\hbar$ in SI units. Of course, the dynamics (\ref{Hdyn}) do emerge from the full quantum theory in the limit of large capacitances; cf.\ Section~\ref{disc}.} The last term in $H$ represents a steep and narrow well, so one expects
\beq
  \phi\equiv\phi_1+\phi_2+\phi_3-\phi_\mathrm{x}
\eeql{def-phi}
to be small and rapidly oscillating, while the other variables can have excursions of order one but move comparatively slowly---ideal for adiabatic elimination of $\phi$. However, from (\ref{Hdyn}) one finds, e.g., $C_1\ddot{\phi}_1=-\phi/L-2eI_{\mathrm{c}1}\sin\phi_1$, so that determination of the $\phi_1$-dynamics to $\Oc(L)$ apparently involves the $\phi$-dynamics to $\Oc(L^2)$, rendering the $L$-expansion quite tedious.

To see what went wrong, consider the rough analogy of a two-dimensional electron gas, where excursions away from the $x$--$y$ plane carry a large penalty in energy. It is intuitively clear that the fast oscillations in the corresponding potential well occur predominantly in the $z$ direction, orthogonal to the easy plane. While the latter plane is uniquely defined by the potential, in our case ``orthogonal'' involves the anisotropic capacitance matrix as well (which thus can be said to induce a metric). It is not difficult to scale the phases such that the charging term is $\propto\vec{Q}^\mathrm{T}\vec{Q}$ (in terms of a \emph{three}-vector $\vec{Q}$; in the quantum case, this yields a charging term $\propto-\nabla^2$), upon which the proper coordinates are found by rotation. However, full orthonormalization turns out to be overkill, and the resulting tedious formulas are hard to interpret; crucial is only that the slow coordinates $\chi,\theta$ are constant along the fast \emph{direction} $(\phi_1,\phi_2,\phi_3)=(C_1^{-1},C_2^{-1},C_3^{-1})$. This is readily achieved; the easiest seems\footnote{The coordinates $\Theta_1,\Theta_2$ (plus $I_m\propto\phi$) in (13) of~\cite{C&O} satisfy the same criterion. They correspond to choosing two zero entries in (\ref{coord-inv}) rather than (\ref{coord}), and are physically equivalent to $\chi,\theta$.}
\begin{gather}
  \begin{pmatrix} \chi \\ \theta \\ \phi{+}\phi_\mathrm{x} \end{pmatrix} =
  \begin{pmatrix} 1 & 0 & -C_3/C_1 \\ 0 & 1 & -C_3/C_2 \\
                  1 & 1 & 1 \end{pmatrix}
  \begin{pmatrix} \phi_1 \\ \phi_2 \\ \phi_3 \end{pmatrix}
  \quad\Rightarrow\label{coord}\\[3mm]
  \begin{pmatrix} \phi_1 \\ \phi_2 \\ \phi_3 \end{pmatrix} =
  C\begin{pmatrix} C_2^{-1}{+}C_3^{-1} & -C_1^{-1} & C_1^{-1} \\[1mm]
                   -C_2^{-1} & C_1^{-1}{+}C_3^{-1} & C_2^{-1} \\[1mm]
                   -C_3^{-1} & -C_3^{-1} & C_3^{-1} \end{pmatrix}
  \begin{pmatrix} \chi \\ \theta \\ \phi{+}\phi_\mathrm{x} \end{pmatrix}\;.
  \label{coord-inv}
\end{gather}
The Kirchhoff equations become
\begin{align}
  \frac{C_1\ddot{\chi}}{2e}&=I_{\mathrm{c}3}\sin\phi_3-I_{\mathrm{c}1}\sin\phi_1\;,
    \label{chi-dyn}\\
  \frac{C_2\ddot{\theta}}{2e}&=I_{\mathrm{c}3}\sin\phi_3-I_{\mathrm{c}2}\sin\phi_2\;,
    \label{theta-dyn}\displaybreak[0]\\
  \frac{1}{2e}\biggl(\ddot{\phi}+\frac{\phi}{LC}\biggr)&=
    -\frac{I_{\mathrm{c}1}}{C_1}\sin\phi_1-\frac{I_{\mathrm{c}2}}{C_2}\sin\phi_2
    -\frac{I_{\mathrm{c}3}}{C_3}\sin\phi_3\;,\label{phi-dyn}
\end{align}
where the $\phi_i$'s on the rhs are simply shorthand for the linear combinations in (\ref{coord-inv}). Indeed, the terms $\sim L^{-1}$ have cancelled in (\ref{chi-dyn}) and (\ref{theta-dyn}). For an interpretation, note that $\ddot{\chi}\propto C_1\dot{Q}_1-C_3\dot{Q}_3$ describes the charging of the \emph{island} between junctions 1 and~3; unlike the charging of, e.g., \emph{capacitor}~1, $\ddot{\chi}$ thus has no net direct contribution from the loop current, and is affected by it only indirectly [in $\Oc(L)$] through the flux-quantization condition.

In these variables, the $L$-expansion to the required order is almost trivial. In $\Oc(L^0)$, one simply sets $\phi=0$ in (\ref{chi-dyn}) and (\ref{theta-dyn}), which is readily verified to yield dynamics equivalent to~(\ref{H3JJ}). In $\Oc(L)$, $\phi$ can be neglected on the rhs of (\ref{phi-dyn}) since it does not occur with a large coefficient. The leading adiabatic solution is then
\beq
  \phi=2eLI^{(0)}(\chi,\theta)\;,
\eeql{adia}
or explicitly $I^{(0)}(\chi,\theta,\phi{=}0)$ in (\ref{I0}). Subsequently, (\ref{adia}) can be substituted into the rhs's of (\ref{chi-dyn}) and~(\ref{theta-dyn}), yielding a self-contained dynamical system with two degrees of freedom.

In preparation for the quantum analysis, it is instructive to also consider a canonical formulation. The Lagrangian reads
\begin{align}
  \mathcal{L}(\chi,\theta,\phi)&=\sum_{i=1}^3\frac{C_i\dot{\phi}_i^2}{8e^2}-V\\
  &=\frac{C}{8e^2}\biggl[\biggl(\frac{C_1}{C_2}{+}\frac{C_1}{C_3}\biggr)\dot{\chi}^2
  -2\dot{\chi}\dot{\theta}+\biggl(\frac{C_2}{C_1}{+}\frac{C_2}{C_3}\biggr)\dot{\theta}^2
  +\dot{\phi}^2\biggr]-V\;,\label{L}
\end{align}
in which one is to substitute (\ref{adia}) for~$\phi$. The transformation (\ref{coord}) has achieved that (\ref{L}) does not contain $\dot{\chi}\dot{\phi}$ or $\dot{\theta}\dot{\phi}$. Such terms, which \emph{do} occur in $\mathcal{L}(\phi_1,\phi_2,\phi)$, would in $\Oc(L)$ result in an extremely unpleasant phase-dependent capacitance matrix in the effective Lagrangian (or Hamiltonian). However, in (\ref{L}) with (\ref{adia}), $\dot{\phi}^2(\chi,\theta)$ is $\Oc(L^2)$, hence negligible for the $(\chi,\theta)$-dynamics. The effective Hamiltonian follows from $Q_\chi=2e\hp\pt\mathcal{L}/\pt\dot{\chi}$ etc., as
\begin{align}
  H(\chi,\theta;Q_\chi,Q_\theta)&\equiv
    \frac{Q_\chi\dot{\chi}+Q_\theta\dot{\theta}}{2e}-\mathcal{L}\\
  &=\frac{1}{2}\biggl[\frac{1}{C_1}\biggl(1{+}\frac{C_3}{C_1}\biggr)Q_\chi^2
    +\frac{2C_3}{C_1C_2}Q_\chi Q_\theta
    +\frac{1}{C_2}\biggl(1{+}\frac{C_3}{C_2}\biggr)Q_\theta^2\biggr]+V\;.\label{H1}
\end{align}
The potential $V$ merits closer inspection: to $\Oc(L)$,
\beq
  V(\chi,\theta)=U_\mathrm{J}(\chi,\theta,\phi{=}0)
    +\left.\frac{\pt U_\mathrm{J}}{\pt\phi}\right|_{\phi{=}0}\phi(\chi,\theta)
    +\frac{\phi^2(\chi,\theta)}{8e^2L}\;.
\eeql{V}
Using (\ref{coord-inv}) to perform the differentiation in the second term explicitly \emph{at constant $\chi,\theta$}, one finds ${\pt_\phi U_\mathrm{J}|}_0=-I^{(0)}(\chi,\theta)/2e$.\footnote{By the same token, one has $I^{(0)}=-2e\hp{\pt\hn H_0/\pt\phi_\mathrm{x}|}_{\chi,\theta}$ for $H_0$ as in~(\ref{H0}). This seemingly obvious relation is simply not true for ${\pt\hn H_\mathrm{3JJ}/\pt\phi_\mathrm{x}|}_{\phi_1,\phi_2}$ in (\ref{H3JJ}), even though $I=-2e\hp{\pt\hn H/\pt\phi_\mathrm{x}|}_{\vec{\phi}}$ does hold between (\ref{H3phi}) and (\ref{I3}) in the three-phase theory. For coupled qubits, one likewise has $\pt\hn H/\pt\hn B_\mathrm{x}=-A_aI_a-A_bI_b$ in (\ref{HM}), (\ref{H6}) below.} Substituting (\ref{adia}) for $\phi(\chi,\theta)$, this means that this term inverts the sign of the third one, yielding
\begin{gather}
  H(\chi,\theta;Q_\chi,Q_\theta)=H_0-\frac{L}{2}I^{(0)}(\chi,\theta)^2\;,\label{Hclass}\\
  \begin{split}
  H_0&=\frac{1}{2}\biggl[\frac{1}{C_1}\biggl(1{+}\frac{C_3}{C_1}\biggr)Q_\chi^2
    +\frac{2C_3}{C_1C_2}Q_\chi Q_\theta
    +\frac{1}{C_2}\biggl(1{+}\frac{C_3}{C_2}\biggr)Q_\theta^2\biggr]\\
    &\quad-E_1\cos\biggl[C\biggl\{\biggl(\frac{1}{C_2}{+}\frac{1}{C_3}\biggr)\chi
        -\frac{\theta}{C_1}+\frac{\phi_\mathrm{x}}{C_1}\biggr\}\biggr]
      -E_2\cos\biggl[C\biggl\{-\frac{\chi}{C_2}
        +\biggl(\frac{1}{C_1}{+}\frac{1}{C_3}\biggr)\theta
        +\frac{\phi_\mathrm{x}}{C_2}\biggr\}\biggr]\\
      &\quad-E_3\cos\biggl[\frac{C}{C_3}\{\phi_\mathrm{x}-\chi-\theta\}\biggr]\;.\\[-8mm]
  \end{split}\label{H0}
\end{gather}
Here, $H_0$ plays the role of $H_\mathrm{3JJ}$ in $(\chi,\theta)$ coordinates. The combination of ``Josephson'' and ``inductive'' corrections in (\ref{Hclass}) would \emph{not} have occurred in, say, $(\phi_1,\phi_2,\phi)$, where thus an awkward potential term would have compensated an awkward charging term [cf.\ below~(\ref{L})]. The sign-flipping of the magnetic term has also been observed around (2) in Ref.~\cite{you}, though not in terms of the detailed current operator $I^{(0)}$ in~(\ref{I0}). Thus, the semiclassical analysis in~\cite{you} seems a valid approximation for stationary currents in the low-lying qubit states.

\setlength{\unitlength}{1mm}
\begin{figure}
   \begin{picture}(75,75)
     \put(0,5){\includegraphics[width=6.5cm]{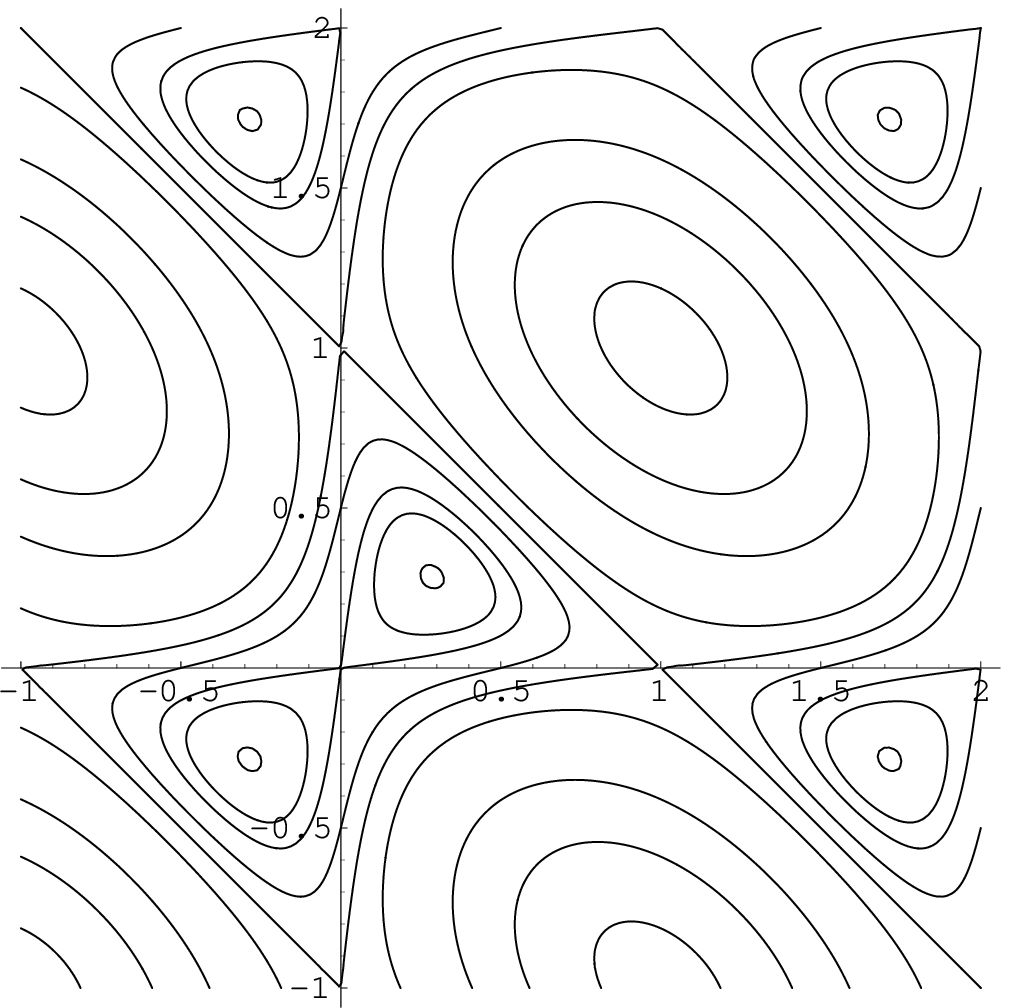}}
     \put(66,27){\footnotesize$\underline{\phi}{}_1/\pi$}
     \put(18,72){\footnotesize$\underline{\phi}{}_2/\pi$}
     \put(33,0){(a)}
   \end{picture}
   \qquad
   \begin{picture}(75,75)
     \put(0,5){\includegraphics[width=6.5cm]{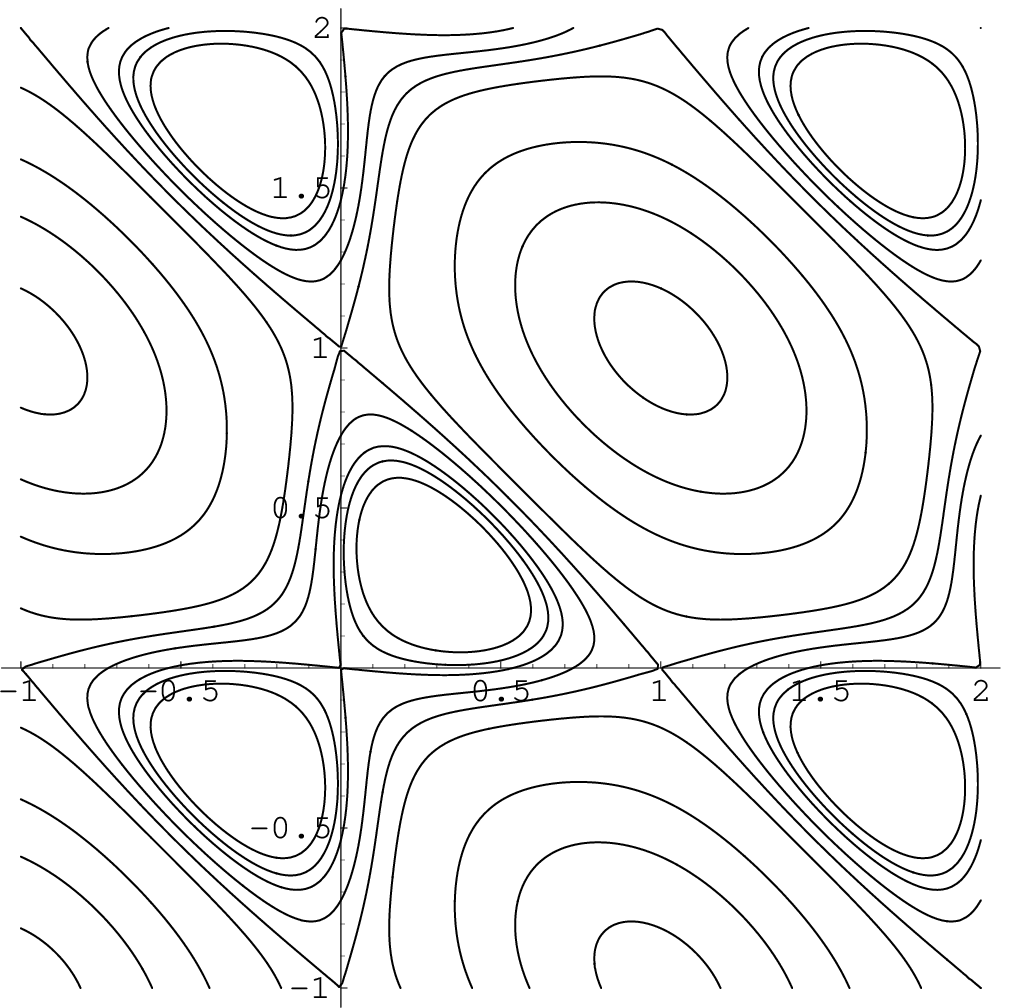}}
     \put(66,27){\footnotesize$\underline{\phi}{}_1/\pi$}
     \put(18,72){\footnotesize$\underline{\phi}{}_2/\pi$}
     \put(33,0){(b)}
   \end{picture}
  \caption{The zeroth-order potential $V_0$ (a) and the first-order one $V$ (b), for $\alpha=0.8$ and $\phi_\mathrm{x}=\pi$. In (b), we also use $e^2LE_1=0.3$. The contours correspond to $V_{(0)}/E_1=-1.42$, $-1.3$, $-1.2$ ($=\alpha{-}2$), $-1$, $-0.8$ ($=-\alpha$), $-0.5$, $0.5$, $1.5$, and $2.5$, the latter encircling the maximum at $\underline{\phi}{}_1=\nobreak\underline{\phi}{}_2=\pi$. The well $V_0(\pm\underline{\phi}^*,\pm\underline{\phi}^*)/E_1=-\alpha-1/2\alpha=-1.425$ at $\underline{\phi}^*=\arccos(1/2\alpha)\approx0.8957$ is shifted to $V(\pm\underline{\xi}^*,\pm\underline{\xi}^*)/E_1\approx-1.7983$ at $\underline{\xi}^*\approx0.9602$. While the shift in well location thus is appreciable, just evaluating the persistent current in the unshifted one, i.e.\ $V(\pm\underline{\phi}^*,\pm\underline{\phi}^*)/E_1=\protect\linebreak -\alpha-1/2\alpha-e^2LE_1(2{-}1/2\alpha^2)\approx-1.7906$, gives a good estimate of the well depth.}
  \label{fig}
\end{figure}

Since $H$ is invariant, one can here and in the following increase the similarity to the conventional treatment by introducing $\underline{\phi}{}_1=C\{(C_2^{-1}{+}C_3^{-1})\chi-\theta/C_1+\phi_\mathrm{x}/C_1\}$ and $\underline{\phi}{}_2=C\{-\chi/C_2+(C_1^{-1}{+}C_3^{-1})\theta+\phi_\mathrm{x}/C_2\}$, so that formally $H_0(\vec{\underline{\phi}}\,)=H_\mathrm{3JJ}(\vec{\phi}\,)$. Since comparison with (\ref{coord-inv}) and (\ref{adia}) shows that $\underline{\phi}{}_{1,2}$ do not coincide with the junction phases to $\Oc(L)$, this formulation is not very useful in derivations. However, in terms of $\vec{\underline{\phi}}$, the double periodicity of~$V$ and any additional symmetries are expressed conveniently, which may be effective numerically [cf.\ below~(\ref{dcSQeff})]. Figure~\ref{fig} shows both the zeroth-order potential $V_0$ and the corrected $V$ of~(\ref{V}), for standard parameters $E_1=\nobreak E_2$, $C_1=C_2$, and $\alpha\equiv E_3/E_1=C_3/C_1=0.8$. In this case $V_0(\underline{\phi}{}_1,\underline{\phi}{}_2)=V_0(\underline{\phi}{}_2,\underline{\phi}{}_1)$, which is preserved in~$V$; since we have chosen a degenerate bias $\phi_\mathrm{x}=\pi$, one also has $V_{(0)}(\underline{\phi}{}_1,\underline{\phi}{}_2)= V_{(0)}(-\underline{\phi}{}_1,-\underline{\phi}{}_2)$. The small self-flux has a clear physical effect: $(I^{(0)})^2=\nabla_{\!\vec{\underline{\phi}}}(I^{(0)})^2=0$ for $\vec{\underline{\phi}}{}^\mathrm{T}\in\{(0,0),(0,\pi),(\pi,0)\}$ by symmetry, so both the location and the height of all saddle-points are unchanged.\footnote{The same holds for the maximum $V_{(0)}(\pi,\pi)/E_1=2{+}\alpha$, but this should have few physical consequences.} Since clearly $-\half L(I^{(0)})^2<0$ in the wells, both the ``easy'' (intra-) and ``hard'' (inter-cell) barriers are increased, and this term's suppression of the tunneling amplitude may not always be negligible.
Finally, the lines $\underline{\phi}{}_1+\underline{\phi}{}_2=\pi\pmod{2\pi}$, which are \emph{straight} equipotentials\footnote{This may be a new observation, and holds for general~$\phi_\mathrm{x}$ even though the saddle-points themselves will move along this line.} through the inter-cell saddle-points of~$V_0$, cease to play this role in~$V$.

\section{Quantum analysis}
\label{quant}

\subsection{Expansion of the Schr\"odinger equation}

In quantum mechanics, one can advantageously use the same variables~(\ref{coord}), avoiding $\pt_\chi\pt_\phi$ and $\pt_\theta\pt_\phi$ in (\ref{Hexp}) below. The large inductive term in (\ref{H3phi}) for $H$ now causes the system to remain in its ground state with respect to $\phi$. To leading order this is a harmonic-oscillator ground state $\psi\sim\e^{-\lm\phi^2}$, and one readily finds
\beq
  \lm=\frac{1}{8e^2}\sqrt{\frac{C}{L}}\;.
\eeql{l}
This unfortunately means that formally $\phi\sim L^{1/4}$ ($\pt_\phi\sim L^{-1/4}$) and strikingly $I\sim L^{-3/4}$. Hence, the expansion to $\Oc(L)$ will turn out to be sixth-order while it was first-order in Section~\ref{class}, and keeping the calculation organized is essential. Since it suffices to study the time-independent problem, the expansion will be analogous to, e.g., fast-variable elimination in (Fokker--Planck) diffusion operators~\cite{opa}.

Using (\ref{Qquant}) and (\ref{coord}) in (\ref{H3phi}), in the eigenvalue equation $(H-E)\psi=0$ we expand
\begin{gather}
  H=-\frac{2e^2}{C}\pt_\phi^2+\frac{\phi^2}{8e^2L}+H_0+H_1\phi+H_2\phi^2+\cdots\;,
    \label{Hexp}\\
  E=E_\mathrm{zp}+E_0+E_1+\cdots\;,\\
  \psi=\psi_0+\psi_2+\cdots\;,\label{psi-exp}
\end{gather}
with the zero-point energy $E_\mathrm{zp}\sim L^{-1/2}$ and $H_0$ as in (\ref{H0}), with $E_i,\psi_i\sim L^{i/4}$, and where we have been able to omit an $L^{-1/4}$ ($L^{1/4}$) term in $E$ ($\psi$) from the outset because $H$ does not have an $L^{-1/4}$ term. One has $H_n=(1/n!){\pt_\phi^n|}_0 U_\mathrm{J}$ for $n\ge1$, and since the calculation below (\ref{V}) is just as valid in quantum as in classical mechanics, $H_1=-I^{(0)}(\chi,\theta)/2e$.

In $\Oc(L^{-1/2})$, one has
\begin{gather}
  \biggl[-\frac{2e^2}{C}\pt_\phi^2+\frac{\phi^2}{8e^2L}-E_\mathrm{zp}\biggr]\psi_0=0
  \quad\Rightarrow\label{zp1}\\
  \psi_0(\chi,\theta,\phi)=\psi_0'(\chi,\theta)\hp\e^{-\lm\phi^2}\;,\qquad
  E_\mathrm{zp}=\frac{1}{2\sqrt{LC}}\label{zp2}\;,
\end{gather}
the obvious answer for zero-point fluctuations of the loop current.

One can use (\ref{zp2}) to factorize the operator in brackets in (\ref{zp1}). In $\Oc(L^0)$, this leads to
\beq
  -\frac{2e^2}{C}\hp\e^{\lm\phi^2}\pt_\phi\bigl[\e^{-2\lm\phi^2}\pt_\phi\bigl(
  \e^{\lm\phi^2}\psi_2\bigr)\bigr]+(H_0{-}E_0)\hp\psi_0'\hp\e^{-\lm\phi^2}\;,
\eeql{OL0}
where one can commute $\e^{-\lm\phi^2}$ through $H_0{-}E_0$. Operating with $\int\!\dr\phi\,\e^{-\lm\phi^2}$, the first term vanishes, so that one obtains the \emph{solvability condition}
\beq
  (H_0{-}E_0)\hp\psi_0'=0\;,
\eeql{L0a}
equivalent to the standard 3JJ theory (\ref{H3JJ}). Substitution back into (\ref{OL0}) renders the latter identical to the leading order, so that
\beq
  \psi_2(\chi,\theta,\phi)=\psi_2'(\chi,\theta)\hp\e^{-\lm\phi^2}\;.
\eeql{psi2}

Proceeding to $\Oc(L^{1/4})$, one has
\beq
  -\frac{2e^2}{C}\hp\e^{\lm\phi^2}\pt_\phi\bigl[\e^{-2\lm\phi^2}\pt_\phi\bigl(
  \e^{\lm\phi^2}\psi_3\bigr)\bigr]
  +\biggl(-\frac{I^{(0)}}{2e}\hp\phi{-}E_1\biggr)\psi_0'\hp\e^{-\lm\phi^2}=0\;,
\eeq
for a solvability condition
\beq
  E_1=0
\eeql{E1}
because the term $\propto I^{(0)}\phi$ cancels by parity. This term does however contribute to $\psi_3$ itself,\footnote{Only solutions bounded in $\phi$ are acceptable.}
\beq
  \psi_3(\chi,\theta,\phi)=
  \biggl[\frac{CI^{(0)}\hp\psi_0'}{16e^3\lm}\hp\phi
  +\psi_3'(\chi,\theta)\biggr]\e^{-\lm\phi^2}\;,
\eeql{psi3}
where the first term in square brackets is readily verified to shift the center of the Gaussian in (\ref{zp2}) to\footnote{Properly, ${\langle\phi\rangle}_{\chi_0,\theta_0}= \langle\phi\hp\delta(\chi{-}\chi_0)\hp\delta(\theta{-}\theta_0)\rangle/ \langle\delta(\chi{-}\chi_0)\hp\delta(\theta{-}\theta_0)\rangle$ is a conditional quantum-mechanical expectation. The vector version below (\ref{psi3vec}) is analogous.} ${\langle\phi\rangle}_{\chi,\theta}$ in accordance with~(\ref{adia}).

In $\Oc(L^{1/2})$, one obtains the first nontrivial solvability condition,\footnote{Of course, one can operate on (\ref{OL2}) with $\int\!\dr\chi\hp\dr\theta\,{(\psi_0')}^*$, yielding $E_2=\langle\psi_0'|(H_2/4\lm)|\psi_0'\rangle/\langle\psi_0'|\psi_0'\rangle$, upon which $\psi_2'$ is determined up to a multiple of $\psi_0'$ (a trivial change in normalization). However, our focus is on deriving the effective theory, not solving it.\label{note}}
\beq
  (H_0{-}E_0)\hp\psi_2'+\biggl(\frac{H_2}{4\lm}{-}E_2\biggr)\psi_0'=0\;,
\eeql{OL2}
where the prefactor is obtained as $\int\!\dr\phi\,\phi^2\hp\e^{-2\lm\phi^2}=(4\lm)^{-1}\!\int\!\dr\phi\,\e^{-2\lm\phi^2}$. While we are only expanding the linear Schr\"odinger equation, it apparently ``knows'' that the proper statistical weight is $|\psi_0|^2$ not $\psi_0$. The correction to $\psi$ follows as
\beq
  \psi_4(\chi,\theta,\phi)=
  \biggl[-\frac{CH_2\hp\psi_0'}{16e^2\lm}\hp\phi^2
  +\psi_4'(\chi,\theta)\biggr]\e^{-\lm\phi^2}\;.
\eeql{psi4}

In $\Oc(L^{3/4})$, the solvability condition reads
\beq
  (H_0{-}E_0)\hp\psi_3'-E_3\psi_0'=0\;,
\eeq
apparently posing the obstacle that both $\psi_3'$ and $E_3$ are unknown in this order. However, one can operate with $\int\!\dr\chi\hp\dr\theta\,{(\psi_0')}^*$ and use~(\ref{L0a}), yielding
\beq
  E_3=0
\eeql{E3}
and $\psi_3'\propto\psi_0'$, so that
\beq
  \psi_3'=0
\eeql{psi3a}
is a convenient choice of normalization. Determination of $\psi_5$ presents no problems, but is not required.

Finally, one arrives at $\Oc(L)$, finding a solvability condition
\beq
  (H_0{-}E_0)\hp\psi_4'-e^2\hn L(H_0{-}E_0)H_2\psi_0'
  -\frac{L}{2}(I^{(0)})^2\psi_0'+\biggl(\frac{H_2}{4\lm}{-}E_2\biggr)\psi_2'
  +\biggl(\frac{12e^4\hn L}{C}H_4{-}E_4\biggr)\psi_0'=0\;.
\eeql{OL4}
What remains is to interpret this in terms of an effective two-phase Schr\"odinger equation.

\subsection{Effective theory}

While odd powers in $L^{1/4}$ were important in the above, most clearly through $I^{(0)}$ in (\ref{psi3}) implementing the quantum counterpart of (\ref{adia}), they cancel in final results such as (\ref{E1}), (\ref{E3}) by $\phi$-parity. What emerges is a two-phase theory involving an expansion in terms of $\sqrt{L}$ in which the zero-point motion (\ref{zp1}) is absent, the leading order being simply
\begin{gather}
  (\tilde{H}_0-\tilde{E}_0)\hp\tilde{\psi}_0=0\;,\\
  \tilde{H}_0=H_0\;,\qquad\tilde{E}_0=E_0\;,\qquad\tilde{\psi}_0=\psi_0'\;.
\end{gather}
Subsequently, (\ref{OL2}) is recognized as the first-order part of
\beq
  [\tilde{H}_0{+}\tilde{H}_1{+}\cdots-(\tilde{E}_0{+}\tilde{E}_1{+}\cdots)]
  (\tilde{\psi}_0{+}\tilde{\psi}_1{+}\cdots)=0\;,
\eeql{eff}
if one sets
\beq
  \tilde{H}_1=\frac{H_2}{4\lm}\;,\qquad\tilde{E}_1=E_2\;,\qquad
  \tilde{\psi}_1=\psi_2'\;.
\eeq

For the next and final relevant order, care is needed in identifying the effective wave function. Since probability should be conserved, $\tilde{\psi}$ must obey $|\tilde{\psi}(\chi,\theta)|^2=\int\!\dr\phi\,|\psi(\chi,\theta,\phi)|^2$, in which $\psi_3$ cancels to $\Oc(L)$ by (\ref{psi3}) and~(\ref{psi3a}). Combining (\ref{psi-exp}), (\ref{zp2}), (\ref{psi2}), and (\ref{psi4}), one has
\beq
  |\psi_0'|^2+2\re[{(\psi_0')}^*\psi_2']+|\psi_2'|^2+2\re[{(\psi_0')}^*\psi_4']
  -2e^2\hn LH_2|\psi_0'|^2+\Oc(L^{3/2})=
  |\tilde{\psi}_0{+}\tilde{\psi}_1{+}\tilde{\psi}_2{+}\cdots|^2\,,
\eeq
which is satisfied by
\beq
  \tilde{\psi}_2=\psi_4'-e^2\hn LH_2\psi_0'\;.
\eeq
Substituting this into (\ref{OL4}), the latter is identified as the second order\footnote{The occurrence of $\tilde{H}_1\tilde{\psi}_1$ in this order represents the effect of $\tilde{H}_1$ in second-order perturbation theory, not a new term in $\tilde{H}$.} of (\ref{eff}), with\footnote{Again, one can solve $\tilde{E}_2=[\langle\tilde{\psi}_0|\tilde{H}_2|\tilde{\psi}_0\rangle+ \langle\tilde{\psi}_1|\tilde{E}_0{-}\tilde{H}_0|\tilde{\psi}_1\rangle]/ \langle\tilde{\psi}_0|\tilde{\psi}_0\rangle$, cf.\ note~\ref{note}. However, in practice direct numerical solution of $\tilde{H}$---though no more accurate---seems preferable over a perturbative approach.\label{again}}
\beq
  \tilde{E}_2=E_4\;,\qquad\tilde{H}_2=\frac{12e^4\hn L}{C}H_4-\frac{L}{2}(I^{(0)})^2\;.
\eeq

Combining the above, one sees that the effective quantum Hamiltonian $\tilde{H}$ again has the form (\ref{Hclass}), but with renormalized Josephson couplings
\beq
  E_i\mapsto E_i'=
  E_i\biggl(1-\frac{e^2\sqrt{LC^3}}{C_i^2}+\frac{e^4\hn LC^3}{2C_i^4}\biggr)\;.
\eeql{renorm}
To this order, the latter is exactly what one would expect from a Gaussian averaging $\tilde{U}_\mathrm{J}(\chi,\theta)= \int\!\dr\phi\,\e^{-2\lm\phi^2}U_\mathrm{J}(\chi,\theta,\phi)/\hn\int\!\dr\phi\,\e^{-2\lm\phi^2}$: the zero-point fluctuations in $\phi$ behave as uncertainty in $\phi_\mathrm{x}$ in (\ref{H0}), which washes out the Josephson potential. The systematic expansion is not prejudiced about ``Josephson'' vs ``inductive'' contributions to the energy, and instead directly provides a term $-\half L(I^{(0)})^2$ to $\tilde{H}$ with the correct sign.

\subsection{Numerics}
\label{numer}

To test the above analysis numerically, introduce the dimensionless inductance $\beta=2\pi LI_\mathrm{c}/\Phi_0=4e^2\hn LE_\mathrm{J}$ ($\Phi_0$ is the flux quantum) and $g=E_\mathrm{J}/\hn E_C$ (with the charging energy~$E_C=e^2\!/2C_1$), and measure all energies in units~$E_\mathrm{J}$. Evaluation of the $L$-expansion is only meaningful for $\beta<1$, and $\beta$ should be varied by at least an order of magnitude. These imply that for some of our parameter values, the energy levels of the full and effective theories will be very close. Hence, we focus on the simplest junction systems (to which the above for the 3JJ qubit is readily adapted), stressing accuracy rather than generality.

First, consider the rf-SQUID, $H=-(4/\hn g)\pt_\phi^2-\cos(\phi{+}\phi_\mathrm{x})+\phi^2\!/2\beta$, for which the effective $\tilde{H}=\sqrt{2/\beta g}-(1{-}\sqrt{\beta/2g}{+}\beta/4g)\cos\phi_\mathrm{x} -\half\beta\sin^2\phi_\mathrm{x}$ is zero-dimensional, i.e.,  a closed expression.\footnote{For convenient comparison, $E_\mathrm{zp}$ (usually an irrelevant constant) is included in~$\tilde{H}$.} To evaluate~$H$, we use that $\phi$ has an effective range $\sim\lm^{-1/2}$ [cf.~(\ref{l})], by truncating the potential with hard walls at $\phi=\pm m\sqrt[4]{\beta/\hn g}$. Discretizing the resulting Dirichlet boundary value problem, the ground-state energy $E$ is found by direct diagonalization, checking for convergence both with respect to the grid spacing $\Delta$ and $m=6,7,8,\ldots$.\footnote{Increasing $m$ too fast, e.g.\ successive doubling, wastes grid points on the wave function's large-$|\phi|$ tail.} Using the known error $\sim\Delta^2$ of lowest-order discretization of $\pt_\phi^2$, one can perform a (highly effective) Richardson extrapolation $E(\Delta{\downarrow}0)\approx[4E(\Delta){-}E(2\Delta)]/3$. The resulting data (not shown) confirm that the effective theory has an error~$\Oc(\beta^{3/2})$.

For a system with a non-trivial effective theory, consider the symmetric dc-SQUID, $H=-(4/\hn g)(\pt_{\phi_1}^2{+}\pt_{\phi_2}^2)-\cos\phi_1-\cos\phi_2 +(\phi_1{+}\phi_2{-}\phi_\mathrm{x})^2\!/2\beta$. With fast and slow variables $\phi\equiv\phi_1{+}\phi_2{-}\phi_\mathrm{x}$, $\chi\equiv\half(\phi_1{-}\phi_2)$ respectively, this becomes
\begin{gather}
  H=-\frac{8}{g}\pt_\phi^2-\frac{2}{g}\pt_\chi^2
  -2\cos\biggl(\frac{\phi{+}\phi_\mathrm{x}}{2}\biggr)\cos\chi+\frac{\phi^2}{2\beta}\;,
  \label{dcSQ}\\
  \tilde{H}=\frac{2}{\sqrt{\beta g}}-\frac{2}{g}\pt_\chi^2
  -2\biggl(1{-}\frac{1}{4}\sqrt{\frac{\beta}{g}}{+}\frac{\beta}{32g}\biggr)
    \cos\biggl(\frac{\phi_\mathrm{x}}{2}\biggr)\cos\chi
  -\frac{\beta}{2}\sin^2\biggl(\frac{\phi_\mathrm{x}}{2}\biggr)\cos^2\chi\;.
  \label{dcSQeff}
\end{gather}
In both (\ref{dcSQ}) and (\ref{dcSQeff}), $2\pi$-periodic boundary conditions are imposed on~$\chi$. This is not the only possible choice, and corresponds to a zero offset charge in the two arms of the SQUID; for large~$g$, the effect of this boundary condition will be small. Observing that both $H$ and $\tilde{H}$ are even in $\chi$, the even sector (containing the ground state\footnote{The ground state tends to be found with the highest accuracy for a given~$\Delta$, but other states are readily found and compared as well.}) can be treated as a Neumann boundary value problem on $0\le\chi\le\pi$. Further details are as for the rf-SQUID.\footnote{The used matrix representation~\cite{NR} makes it convenient to allocate only half of a symmetric matrix in a way that is transparent to the rest of the program, as long as the latter never accesses matrix elements above the main diagonal.}

\begin{table}
\begin{tabular}{|c|c|c|r|r|r|} \hline
$g$ & $\phi_\mathrm{x}$ & $\beta$ & $E$~~~~~\, & $E_0$~~~~\; & $\tilde{E}$~~~~~\, \\
\hline\hline 40 & $\pi$ & 0.32 & $0.471441$ & 0.559017 & 0.463997 \\
\hline 40 & $\pi$ & 0.08 & $1.097919$ & 1.118034 & 1.097038 \\
\hline 40 & $\pi$ & 0.02 & $2.231116$ & 2.236068 & 2.231006 \\
\hline\hline 40 & $\pi/4$ & 0.32 & $-1.057313$ & $-1.076987$ & $-1.059445$ \\
\hline 40 & $\pi/4$ & 0.08 & $-0.503724$ & $-0.517970$ & $-0.503857$ \\
\hline 40 & $\pi/4$ & 0.02 & $0.608463$ & 0.600064 & 0.608456 \\
\hline\hline 1 & $\pi/4$ & 0.32 & $2.998577$ & 2.870307 & 2.996471 \\
\hline 1 & $\pi/4$ & 0.08 & $6.476253$ & 6.405841 & 6.476015 \\
\hline 1 & $\pi/4$ & 0.02 & $13.513831$ & 13.476908 & 13.513799 \\
\hline
\end{tabular}
\caption{The ground-state energy $E$ of a symmetric dc-SQUID, compared to the standard estimate $E_0$ which ignores inductance effects entirely, and to the prediction of the effective theory $\tilde{E}$, accounting for renormalization of the Josephson coupling and for self-flux. Special attention has been paid at $g=40$, $\phi_\mathrm{x}=\pi/4$, $\beta=0.02$ to obtain the very small $\tilde{E}-E$ reliably. For $\phi_\mathrm{x}=\pi$, the zeroth-order problem reduces to a free rotor, hence $E_0=2/\hn\sqrt{\beta g}$.}
\label{data}
\end{table}

Some results are shown in Table~\ref{data}, together with the standard zeroth-order prediction~$E_0$ which results by dropping all correction terms in~(\ref{dcSQeff}). In all cases, reducing $\beta$ by a factor~4 reduces the error $\tilde{E}-E$ by at least a factor~$\sim$8, as predicted. We note without explanation that for $g=40$, $\phi_\mathrm{x}=\pi/4$, this factor seems to be rather $\sim$16, as if the next $\Oc(\beta^{3/2})$ correction were very small numerically or cancelled altogether. In contrast, $E_0-E$ only goes down by a factor~$\sim$2, except for $\phi_\mathrm{x}=\pi$, in which case the effective theory has a zero Josephson coupling so that its $\Oc(\sqrt{\beta})$ renormalization does not matter. Still, even in the latter case the self-flux term is nontrivial, so that an $\Oc(\beta)$ error renders $E_0$ less accurate than~$\tilde{E}$. The last three rows emphasize that, in spite of this paper's title, our $L$-expansion is not at all limited to the ``flux-qubit'' regime $g\gg1$. This is useful, since small\nobreakdash-$L$, small\nobreakdash-$C$ dc-SQUIDs are commonly used as tunable compound junctions, and 3JJ qubits in an intermediate-$g$ regime have also been considered recently~\cite{amin}.

In summary, the transparent effective Hamiltonian $\tilde{H}$ can be used with excellent accuracy also in many cases where the zeroth-order $H_0$ would be unsatisfactory. In the actual computations, the advantages in speed and storage requirements are evident; these will persist even when less naive algorithms are employed in the study of more complex devices.

\subsection{Comparison}

Inductance effects in flux qubits have previously been studied in Ref.~\cite{C&O},\footnote{In that paper's opening paragraph, $2\pi I_\mathrm{c}/\Phi_0$ defines $L_\mathrm{J}^{-1}$ not $L_\mathrm{J}^{\nl}$. The order estimates below (1) give the number of grid points, while the ``computational time'' grows faster. Equation~(3) and the one right above have a sign problem when compared to (2) and Fig.~1(a). A constant $2E_\mathrm{J}$ is lost between (6)--(7) and~(9). In the definition of $\tilde{\Theta}_m$ below (14), the factor~2 in the second term is spurious. In the expression for $\hbar\omega_0$ below (16), $E_c/\hn E_\mathrm{J}$ should read $E_cE_\mathrm{J}$. The figures are irreproducible since $E_\mathrm{J}/\hn E_c$ is not given.} with the examples of loops containing one through three junctions. For one junction, the rf-SQUID, the two treatments are still largely equivalent, and agree on the $\Oc(\sqrt{L})$ contribution to the energy [due to the second term in~(\ref{renorm})]; this presumably is the main reason for the improved agreement with numerics in their Fig.~1(b). In $\Oc(L)$, the last term in (5) gives the self-flux contribution $-\half L(I^{(0)})^2$; however, in the same order the renormalization effect [third term in (\ref{renorm})] is missing, because in the transition from (3) to (4), the Josephson potential was expanded to second instead of to the required fourth order in~$I$.

For loops with multiple junctions, several problems arise in the treatment of~\cite{C&O}. First, the physical role of the loop's series capacitance is not identified,\footnote{In general, not taking independent capacitances and Josephson couplings makes it more difficult to conceptually separate the effects of the two.} as is most clearly seen in the third term of (9), written in terms of the parallel capacitance $2C$ (the appropriate effective mass for the $\Theta$ coordinate) while this term can be naturally expressed as $\half(C\hn/2)L^2\dot{I}^2$ [cf.~(\ref{dcSQ})]. Observing that the series capacitance of the 3JJ qubit is $[\alpha/(1{+}2\alpha)]C$ would have uncovered the missing factor $\half$ in the third term of (14), which presently carries through to underestimating $\omega_0$ below (16) by a factor $\sqrt{2}$---which is not detected in Fig.~3(b) because the latter does not feature three-phase numerics.

Second, the renormalization effect is not estimated correctly. The last term of~(10) is neglected because it involves $L^2$, which is small, without checking whether $\tilde{I}^2$ could possibly be large. In fact, this term's counterpart in (4) yields the dominant correction in~(5). For three junctions, the corresponding term is simply omitted from~(15). Note that Fig.~2(b) is given on such a scale, and for such extremely small $\beta$, that the apparent agreement with numerics only verifies the zeroth-order result which never was in doubt.

Third and most serious, the analysis is conceptually incorrect in averaging the dynamics of the fast variable over the dynamics of the slow ones [first below~(9)] instead of the other way round. In the final results (11) and (16), this leads (in our notation) to a perturbative energy correction $-\half L\langle I^{(0)}\rangle^2$ instead of the proper $-\half L\langle(I^{(0)})^2\rangle$ (cf.\ note~\ref{again}). This is inconsistent with the classical limit of Section~\ref{class} and underestimates the magnitude of the self-flux effect---most dramatically for the two lowest eigenstates of the degenerately biased 3JJ qubit (cf.~our Fig.~\ref{fig}), where it would incorrectly predict this effect to vanish entirely. See also the remark on the coupled case in Section~\ref{disc} below.

\section{Two coupled qubits}
\label{couple}

Consider two SQUIDs $a$ and $b$ side by side. The directions of the fluxes have to be chosen consistently in each loop w.r.t.\ the external field, and their magnitude is given as
\beq
  \bm{\Phi}-\bm{\Phi}_\mathrm{x}=\mathbb{L}\bm{I}\;,
\eeql{PhiL}
where $\bm{\Phi}=(\Phi_a,\Phi_b)^{\!\mathrm{T}}$ etc., and where
\beq
  \mathbb{L}=\begin{pmatrix} \M L_a & -M \\ -M & \M L_b \end{pmatrix}
\eeql{Lmat}
so that the antiferromagnetic coupling is characterized by a positive~$M$. We can assume a homogeneous external field, so that the applied fluxes $\bm{\Phi}_\mathrm{x}=(A_a,A_b)^{\!\mathrm{T}}B_\mathrm{x}$ are proportional to the loop areas $A_{a,b}$. The magnetic energy can now be written as\footnote{Expanding the bilinear form in (\ref{HM-I}), the \emph{anti}ferromagnetic interaction is seen to involve a term $-MI_aI_b$, in contrast to the positive last term in (\ref{HM}). This is counterintuitive, but a simple consequence of $2\times2$ matrix inversion. It may be clarifying to compare (anti)ferromagnetic configurations \emph{at constant absolute fluxes in the loops}. Due to the mutual inductance, antiparallel fluxes can be generated with smaller $|I|$'s, reducing the value of $\half L_a^{\protect\vphantom{2}}I_a^2$ etc., which one would naively discard as a small part of the ``free'' Hamiltonian. That is, while total energy is conserved, the designation of an ``interaction'' part can be somewhat arbitrary, so it is best to be as consistent as possible and retain all terms in~(\ref{HM-I}). The electrostatic counterpart may be more familiar in the field, cf.\ the remark below~(\ref{HM}).}
\begin{align}
  H_\mathrm{M}&=\half\bm{I}^\mathrm{T}\mathbb{L}\bm{I}\label{HM-I}\\
  &=\half(\bm{\Phi}{-}\bm{\Phi}_\mathrm{x})^{\!\mathrm{T}}\mathbb{L}^{\!-1}
    (\bm{\Phi}{-}\bm{\Phi}_\mathrm{x})\notag\\
  &=\frac{(\Phi_a{-}A_aB_\mathrm{x})^2}{2L_a(1{-}k^2)}
    +\frac{(\Phi_b{-}A_bB_\mathrm{x})^2}{2L_b(1{-}k^2)}
    +\frac{M(\Phi_a{-}A_aB_\mathrm{x})(\Phi_b{-}A_bB_\mathrm{x})}{L_aL_b(1{-}k^2)}\;,
    \label{HM}
\end{align}
where $k^2\equiv M^2\!/\hn L_aL_b\ll1$ will never (need to) be assumed. The factors $(1{-}k^2)^{-1}$~\cite{ralph} are often ignored, but should not be surprising given the nontrivial effective capacitances one is used to seeing in charge-qubit Hamiltonians. Physically, they arise since, e.g., $\Phi_a-A_aB_\mathrm{x}$ is not simply the self-flux of loop~$a$, but includes a mutual contribution.\footnote{For $k\ua1$, (\ref{HM}) seems to diverge. However, physically this limit can only be achieved with two loops right on top of each other. The resulting \emph{ferro}magnetic interaction is described by $M<0$ in (\ref{Lmat}), so that the last term in (\ref{HM}) tends to cancel the other two. Indeed, we have merely rewritten the finite (\ref{HM-I}).}

Two coupled 3JJ qubits can now be described by\footnote{Not all of our assumptions (cf.\ note~\ref{series}) may hold in a figure-eight geometry~\cite{paauw} if part of the inductance is distributed along the shared leg. Therefore, this case deserves special attention.\label{8}}
\beq
  H=\sum_{c=a,b}\sum_{i=1}^3\left[\frac{Q_{ic}^2}{2C_{ic}}-E_{ic}\cos\phi_{ic}\right]
  +H_\mathrm{M}\;,
\eeql{H6}
with $\Phi_c=\sum_{i=1}^3\phi_{ic}/2e$ in (\ref{HM}). This description is valid for arbitrary inductances; the price to pay is that (\ref{H6}) involves six degrees of freedom. Let us turn to the quantum expansion right away, limiting ourselves to outlining the proper generalization of Section~\ref{quant}.

In $\Oc(L^{-1/2})$, the anisotropic oscillator problem is solved by
\beq
  \psi_0=\psi_0'(\chi_a,\theta_a;\chi_b,\theta_b)\hp
  \e^{-\bm{\phi}^\mathrm{T}\!\Lm\bm{\phi}}\;;
\eeql{psi60}
the inverse covariance matrix $\Lm$ is found from $\Lm\mathbb{C}^{-1}\Lm=\mathbb{L}^{-1}\!/64e^4$, with $\mathbb{C}\equiv\diag(C_a,C_b)$ diagonal since the coupling between the qubits is purely inductive. One obtains
\beq
  \Lm=\frac{1}{8e^2}\sqrt{\mathbb{C}}
  \bigl[\sqrt{\mathbb{C}}\mathbb{L}\sqrt{\mathbb{C}}\bigr]^{-1/2}
  \sqrt{\mathbb{C}}\;,\qquad
  E_\mathrm{zp}=4e^2\biggl(\frac{\Lm_{aa}}{C_a}+\frac{\Lm_{bb}}{C_b}\biggr)\;.
\eeql{Lambda}
Since even $2\times2$ matrix square roots are slightly tedious, we leave $\Lm$ unevaluated in most formulas [cf.\ (\ref{Laa}) below]. The correlation (entanglement) expressed by the Gaussian in (\ref{psi60}) seems physically reasonable: for, e.g., again two small loops on top of each other, deviations of $\Phi_{a,b}$ from each other should be even more strongly suppressed than deviations from $\Phi_\mathrm{x}$.

In $\Oc(L^0)$, one has
\beq
  -2e^2\hp\e^{\bm{\phi}^\mathrm{T}\!\Lm\bm{\phi}}\hp\nabla_{\bm{\phi}}^\mathrm{T}\hp
  \e^{-2\bm{\phi}^\mathrm{T}\!\Lm\bm{\phi}}\hp\mathbb{C}^{-1}\nabla_{\bm{\phi}}\hp
  \e^{\bm{\phi}^\mathrm{T}\!\Lm\bm{\phi}}\psi_2+\e^{-\bm{\phi}^\mathrm{T}\!\Lm\bm{\phi}}
  (H_0{-}E_0)\hp\psi_0'=0\;.
\eeq
Since $\nabla_{\bm{\phi}}^\mathrm{T}$ has two components, one needs a double integral $\int\!\dr\phi_a\hp\dr\phi_b\,\e^{-\bm{\phi}^\mathrm{T}\!\Lm\bm{\phi}}$ to cancel it, yielding a single solvability condition formally equal to~(\ref{L0a}). However, in the present case we additionally observe the decoupling
\beq
  H_0=H_{0a}+H_{0b}\qquad\Rightarrow\qquad E_0=E_{0a}+E_{0b}\;,\quad
  \psi_0'=\psi_{0a}'(\chi_a,\theta_a)\hp\psi_{0b}'(\chi_b,\theta_b)\;.
\eeq
Thus, the leading six-phase wave function $\psi_0$ can be strongly entangled, while the leading four-phase one $\psi_0'$ factorizes. Subsequently, $\psi_2$ is obtained analogously to~(\ref{psi60}).

In $\Oc(L^{1/4})$, one again finds $E_1=0$, while
\beq
  \psi_3=\biggl[\bm{\phi}^\mathrm{T}
  \frac{\mathbb{C}\Lm^{-1}\bm{I}^{(0)}}{16e^3}\hp\psi_{0a}'\psi_{0b}'
  +\psi_3'(\chi_a,\theta_a;\chi_b,\theta_b)\biggr]
  \e^{-\bm{\phi}^\mathrm{T}\!\Lm\bm{\phi}}\;.
\eeql{psi3vec}
A Gaussian vector integral, followed by use of (\ref{Lambda}), confirms the expected $\langle\bm{\phi}\rangle=2e\mathbb{L}\bm{I}^{(0)}$.

In $\Oc(L^{1/2})$, the correction to $H$ takes the form $H_{2a}^{\vphantom{2}}\phi_a^2+H_{2b}^{\vphantom{2}}\phi_b^2$ since (\ref{H6}) has no Josephson interaction [cf.\ the remark below~(\ref{psi60})]. This does \emph{not} generate any interaction when $\bm{\phi}$ is integrated out with the entangled weight $\e^{-2\bm{\phi}^\mathrm{T}\!\Lm\bm{\phi}}$, so that
\begin{gather}
  \psi_2'=\psi_{2a}'(\chi_a,\theta_a)\hp\psi_{2b}'(\chi_b,\theta_b)\;,\qquad
  E_2=E_{2a}+E_{2b}\;,\\
  (H_{0a}{-}E_{0a})\psi_{2a}'
  +[{\textstyle\frac{1}{4}}(\Lm^{-1})_{aa}H_{2a}{-}E_{2a}]\psi_{0a}'=0
\end{gather}
and similarly for $a\leftrightarrow b$. One could proceed to solve for $\psi_4$ but this will be omitted, since all that really matters is its relation to the effective wave function $\tilde{\psi}_2=\int\!\dr\phi_a\hp\dr\phi_b\, \psi_4\hp\e^{-\bm{\phi}^\mathrm{T}\!\Lm\bm{\phi}}/\!\int\!\dr\phi_a\hp\dr\phi_b\, \e^{-2\bm{\phi}^\mathrm{T}\!\Lm\bm{\phi}}$.

The next order $\Oc(L^{3/4})$ again is comparatively uninteresting, as it mainly justifies taking $E_3=\psi_3'=0$.

Finally, in $\Oc(L)$, everything can be combined, and the effective Hamiltonian can be read off. Doing so leads to the central result of this paper,
\beq
  \tilde{H}=H_{0a}'+H_{0b}'-\half (\bm{I}^{(0)})^\mathrm{T}\mathbb{L}\bm{I}^{(0)}\;,
\eeql{Hfinal}
where $H_{0a}'$, $H_{0b}'$ are as in (\ref{H0}) (with $\phi_{\mathrm{x}c}=2eA_cB_\mathrm{x}$), except for the renormalizations
\beq
  E_{ic}\mapsto E_{ic}'=E_{ic}\biggl[1
  -\biggl(\frac{C_c}{C_{ic}}\biggr)^{\!\!2}\frac{(\Lm^{-1})_{cc}}{8}
  +\biggl(\frac{C_c}{C_{ic}}\biggr)^{\!\!4}\frac{(\Lm^{-1})_{cc}^2}{128}\biggr]
  \qquad(c=a,b)\;.
\eeq

\section{Discussion}
\label{disc}

One can view (\ref{Hfinal}) in at least two ways. Experimentally, the bare couplings $E_{ic}$ will usually be unknown and the diagonal contributions from the last term may be comparatively small. In that case, most interesting will be the interaction $MI^{(0)}_a(\chi_a,\theta_a)\hp I^{(0)}_b(\chi_b,\theta_b)$, derived with the antiferromagnetic sign without any remaining ambiguity. Also, this interaction is given in terms of explicitly known current operators, so that in e.g.\ the two-level (qubit) approximation on has $H_\mathrm{int}=M\lvert I_a\rvert\lvert I_b\rvert\hp\sigma^z_a\sigma^z_b$ in the flux basis without additional handwaving. Contrast~\cite{C&O}, where improper averaging of the $I^{(0)}$'s would have led to $\langle H\rangle=\cdots+M\langle I_a^{(0)}\rangle\langle I_b^{(0)}\rangle$, i.e., no interaction between the qubits.

On the other hand, for detailed numerical comparison between the four- and six-phase theories, every last bit of parameter dependence needs to be made explicit. Therefore, we take the square root in (\ref{Lambda}) in the eigenbasis, yielding
\begin{gather}
  \begin{split}(\Lm^{-1})_{aa}=\frac{2\sqrt{2}e^2}{C_a}\biggl[&
  \biggl(1+\frac{L_aC_a{-}L_bC_b}{\Omega}\biggr)\sqrt{L_aC_a{+}L_bC_b{+}\Omega}\\
  &+\biggl(1+\frac{L_bC_b{-}L_aC_a}{\Omega}\biggr)\sqrt{L_aC_a{+}L_bC_b{-}\Omega}
  \biggr]\;,\end{split}\label{Laa}\\
  \Omega=\sqrt{(L_aC_a{-}L_bC_b)^2+4M^2C_aC_b}\;,\\
  \det\Lm=\frac{1}{64e^4}\sqrt{\frac{C_aC_b}{L_aL_b(1{-}k^2)}}\;,
\end{gather}
while $(\Lm^{-1})_{bb}$ follows by $a\leftrightarrow b$. The second square root in (\ref{Laa}) has a positive argument since $k^2<1$ (positive-definiteness of~$\mathbb{L}$). Note that (\ref{Laa}) reduces to $\lm_a^{-1}$ as in (\ref{l}) for $\Omega\To\pm(L_aC_a-L_bC_b)$ (i.e., regardless of which of the uncoupled qubits has the larger $LC$).

As already seen in Section~\ref{numer}, the $L$-expansion involves two different dimensionless parameters---as expected in a system with Coulomb, Josephson, and magnetic energies. ``Classical'' self-flux effects go as $LI^2\!/E_1\sim\beta$. On the other hand, from (\ref{renorm}) one reads off that the ``quantum'' zero-point effects go as $e^2\sqrt{L/C}\sim\lm^{-1}\sim E_C/\omega_{LC}$. Thus, the former (latter) parameter does not contain the charging (Josephson) energy, which is also the reason why, to the given order, their contributions are separated in the final results. In particular, for comparatively large junctions and not too small area, the $\Oc(L)$ self-flux effect could dominate over the $\Oc(\sqrt{L})$ renormalization within the domain of validity of our expansion---as was implicitly used in Fig.~\ref{fig}. However, in the experiments for which all parameters are available, one has $(LI_\mathrm{c}/\Phi_0,e^2\sqrt{L/C})=(0.003,0.015)$~\cite{VdW} and $(0.007,0.02)$~\cite{Rabi,IMT} respectively, so that, if anything, the renormalization effect dominates.

Extension to $>2$ qubits, using the same vector notation, is immediate. It is hoped that the methods presented will have additional use in other devices with near-constrained variables.

I thank M.H.S. Amin (who also pointed out the exception in note~\ref{8}), M.~Grajcar, J.F. Ralph, G.~Rose, M.F.H. Steininger, M.C. Thom, J.Q. You, and A.M. Zagoskin for fruitful discussions and comments on the manuscript.

\end{document}